\documentclass{ws-procs9x6}

\begin{document}

\title{The $r$-Process: Current Understanding and Future 
Tests\footnote{\uppercase{T}his work is supported in part by the
\uppercase{US} \uppercase{D}epartment of \uppercase{E}nergy under
grants \uppercase{DE}-\uppercase{FG}02-87\uppercase{ER}40328 
and \uppercase{DE}-\uppercase{FG}02-00\uppercase{ER}41149.}}

\author{Yong-Zhong Qian}

\address{School of Physics and Astronomy, \\
University of Minnesota, \\ 
Minneapolis, MN 55455, USA\\ 
E-mail: qian@physics.umn.edu}

\maketitle

\abstracts{
Current understanding of the $r$-process is summarized in terms of 
the astrophysical sites, the yield patterns, and the role of neutrinos.
The importance of observational and experimental tests is 
emphasized. A number of future tests regarding
the above three aspects of the $r$-process are discussed.}

\section{Introduction}
The goal of this contribution is to present a summary of current 
understanding of the $r$-process and to suggest a number of 
observational and experimental tests that can lead to further progress.
Three aspects of the $r$-process will be discussed: the 
astrophysical sites, the yield patterns, and the role of neutrinos.
A more detailed review of recent progress in understanding the
$r$-process can be found in [\refcite{qian03}].

\section{Astrophysical Sites}
Despite decades of studies, we still do not have a self-consistent
model that can produce the conditions for $r$-process nucleosynthesis.
On the other hand, two major categories of candidate sites have been
proposed: core-collapse supernovae (e.g. 
[\refcite{wb92}--\refcite{wo94}]) and neutron star mergers (e.g. 
[\refcite{la74}--\refcite{frt99}]). A simple argument was made in 
[\refcite{qian00}] to favor core-collapse supernovae over neutron star 
mergers as the major site for the $r$-process based on observations
of abundances in metal-poor Galactic halo stars. This argument was
borne out by a detailed numerical study on the chemical evolution of 
the early Galaxy [\refcite{ar04}], which calculated $r$-process abundances
in stars for both core-collapse supernova and neutron star merger models
and compared them with data. The basic conclusions from
these studies are that observations of metal-poor stars are consistent 
with $r$-process enrichment by core-collapse supernovae (see also 
[\refcite{is99}]) but are in conflict with neutron star mergers
being the major $r$-process site. The conclusion regarding neutron star 
mergers is quite robust so long as the Galactic rate of these events
is much lower than that of core-collapse supernovae. For example, with
a typical Galactic rate of $\sim (100\ {\rm yr})^{-1}$ for core-collapse 
supernovae and of $\sim (10^5\ {\rm yr})^{-1}$ for neutron star mergers,
the Fe enrichment of the interstellar medium (ISM) by core-collapse 
supernovae would be much more frequent than the $r$-process enrichment
by neutron star mergers if the latter were the major $r$-process site.
In this case, stars with very low Fe abundances formed from an ISM where 
only a few core-collapse supernovae had occurred would have received no
contributions from neutron star mergers and therefore, would have no 
$r$-process elements such as Eu. This is in strong conflict with the 
substantial Eu abundances observed in stars having Fe abundances as low as 
$\sim 10^{-3}$ times solar and also with the substantial Ba abundances 
observed in stars having Fe abundances as low as $\sim 10^{-4}$ times 
solar\footnote{At such low metallicities corresponding to early times, 
only the $r$-process associated with fast-evolving massive progenitors 
can contribute to the Ba in the ISM.} (e.g. [\refcite{mc95}]). 
These observations can be accounted for only if
neutron star mergers could make major $r$-process contributions at
a rate close to that of Fe enrichment by core-collapse supernovae.
Such a high neutron star merger rate is very unlikely and it is much more
probable that core-collapse supernovae are responsible for both the 
$r$-process and Fe enrichment of metal-poor stars. A firm upper limit
on the rate of neutron star mergers may be provided by LIGO that is being
built to detect gravitational wave signals from such events. In the rest
of the discussion, it will be assumed that core-collapse supernovae are
the major $r$-process site.

\subsection{Diverse $r$-Process Sources}
Assuming that core-collapse supernovae are sources of the $r$-process 
elements and Fe, we still have to answer the following questions:
(1) Does $r$-process production vary from event to event?
(2) How does $r$-process production correlate with Fe production?
(3) How does $r$-process and Fe production depend on the progenitors
of core-collapse supernovae? These questions will be addressed mostly
from the observational side below.

Several observations support that there are at least two distinct kinds
of $r$-process sources. Meteoritic data on extinct radioactivities in
the early solar system indicate that at least some $r$-process events
produce $^{182}$Hf but not $^{129}$I (e.g. [\refcite{wa96}]). In addition,
observations of the metal-poor star CS 22892-052 [\refcite{sn00}] show that 
while the elements Ba and above with mass numbers $A>130$ closely follow 
the solar $r$-process pattern, the elements Rh and Ag with $A<130$ fall 
below the extension of this pattern to the region of $A<130$ 
(see Fig. 1). The $r$-process pattern in this star is rather close to 
that in another metal-poor star CS 31082-001 [\refcite{hi02}] (see Fig. 4). 
Both the meteoritic data and observations of metal-poor stars indicate that
in order to obtain the overall solar $r$-process pattern, there should
be a source producing mainly the heavy $r$-process nuclei with
$A>130$ and another producing mainly the light $r$-process nuclei with
$A\leq 130$.

\begin{figure}[htb]
\begin{center}
\includegraphics[scale=0.4,angle=270]{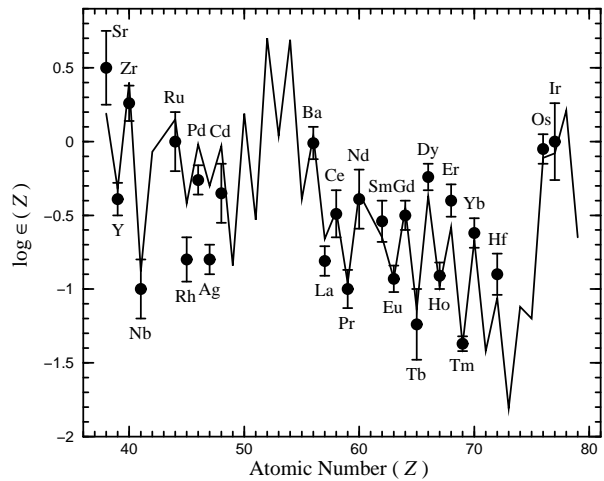}
\caption{The observed abundances in CS 22892-052 (filled circles with 
error bars: [\protect\refcite{sn00}]) compared with the solar $r$-process 
pattern (solid curve: [\protect\refcite{ar99}]) that is translated to pass 
through the Eu data. The data on the heavy $r$-process elements from Ba to 
Ir are in excellent agreement with the translated solar $r$-process pattern. 
However, the data on the light $r$-process elements Rh and Ag clearly fall 
below this pattern. The abundance of element E is given in the spectroscopic 
notation $\log\epsilon({\rm E})\equiv\log({\rm E/H})+12$, where E/H is
the number ratio of E to H atoms in the star.}
\end{center}
\end{figure}

The relation between production of the heavy $r$-process nuclei and that
of Fe can be inferred by comparing the abundances in the 
metal-poor stars CS 31082-001 [\refcite{hi02}], HD 115444, and HD 122563 
[\refcite{we00}] (see Fig. 2). While the heavy $r$-process elements closely 
follow the solar $r$-process pattern for all three stars, the absolute 
abundances of these elements differ by a factor of $\sim 100$. However, 
the absolute abundances of the elements from O to Ge with $A<75$ are
approximately the same for these stars. This clearly demonstrates that
the source for the heavy $r$-process nuclei in these stars cannot produce 
any of the elements from O to Ge including Fe (e.g. [\refcite{qw02}]). The 
core-collapse supernovae associated with this source may have progenitors 
of $\sim 8$--$10\,M_\odot$, which develop O-Ne-Mg cores with very thin 
shells before collapse [\refcite{no84,no87}]. The supernova shock produces 
essentially no nucleosynthesis as it propagates through the thin shells. 
So no elements from O to Ge are made. However, production of the heavy 
$r$-process nuclei could occur in the material ejected from the 
newly-formed neutron star (e.g. [\refcite{wa03}]).
This can then explain why these nuclei are not produced together with
the elements from O to Ge. In addition, the source for the heavy 
$r$-process nuclei in the stars shown in Fig. 2
can also be associated with accretion-induced collapse
(AIC) of a white dwarf into a neutron star in binaries [\refcite{qw03}]
as no elements from O to Ge are produced in this case, either.

\begin{figure}[htb]
\begin{center}
\includegraphics[scale=0.34]{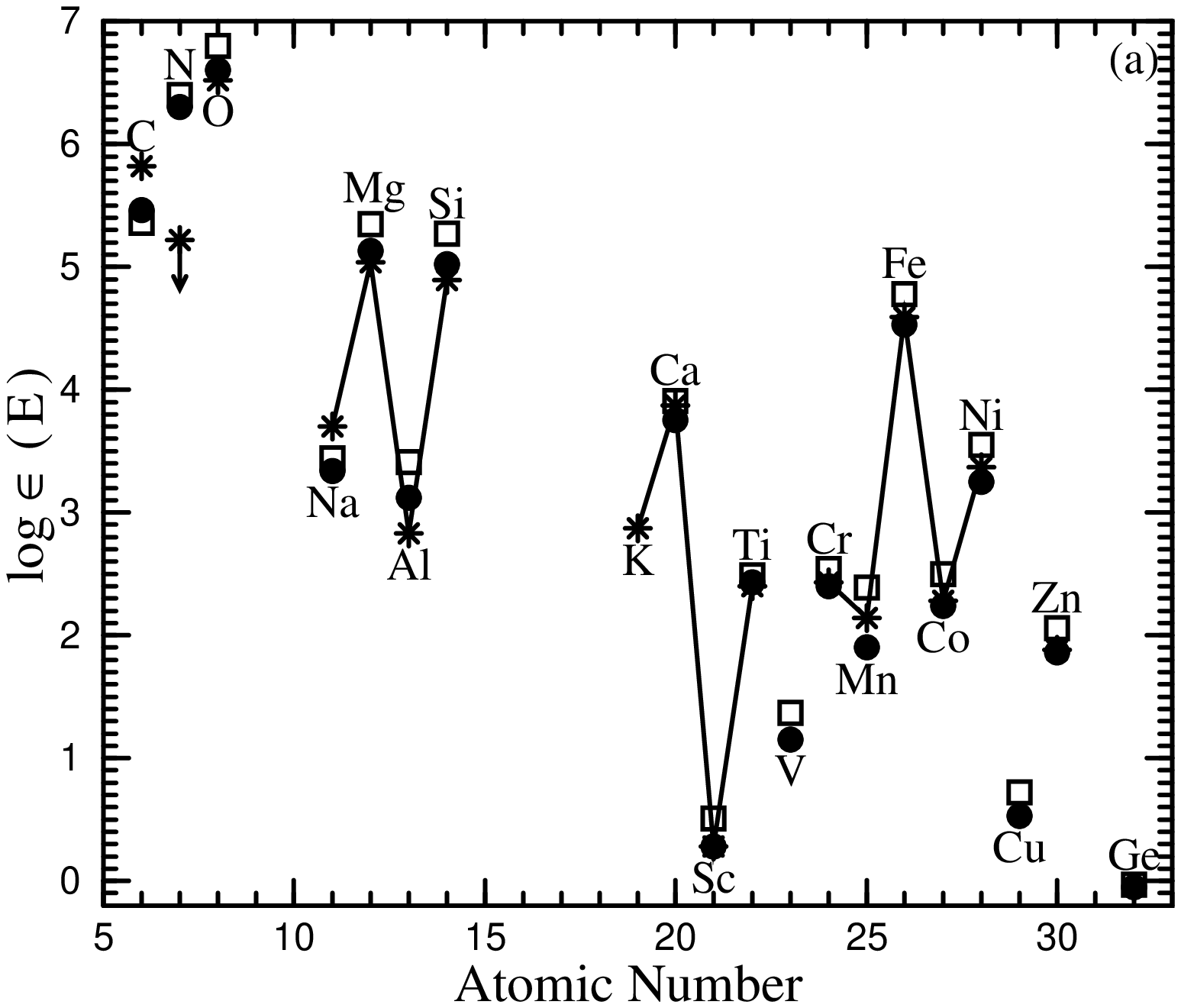}
\includegraphics[scale=0.34]{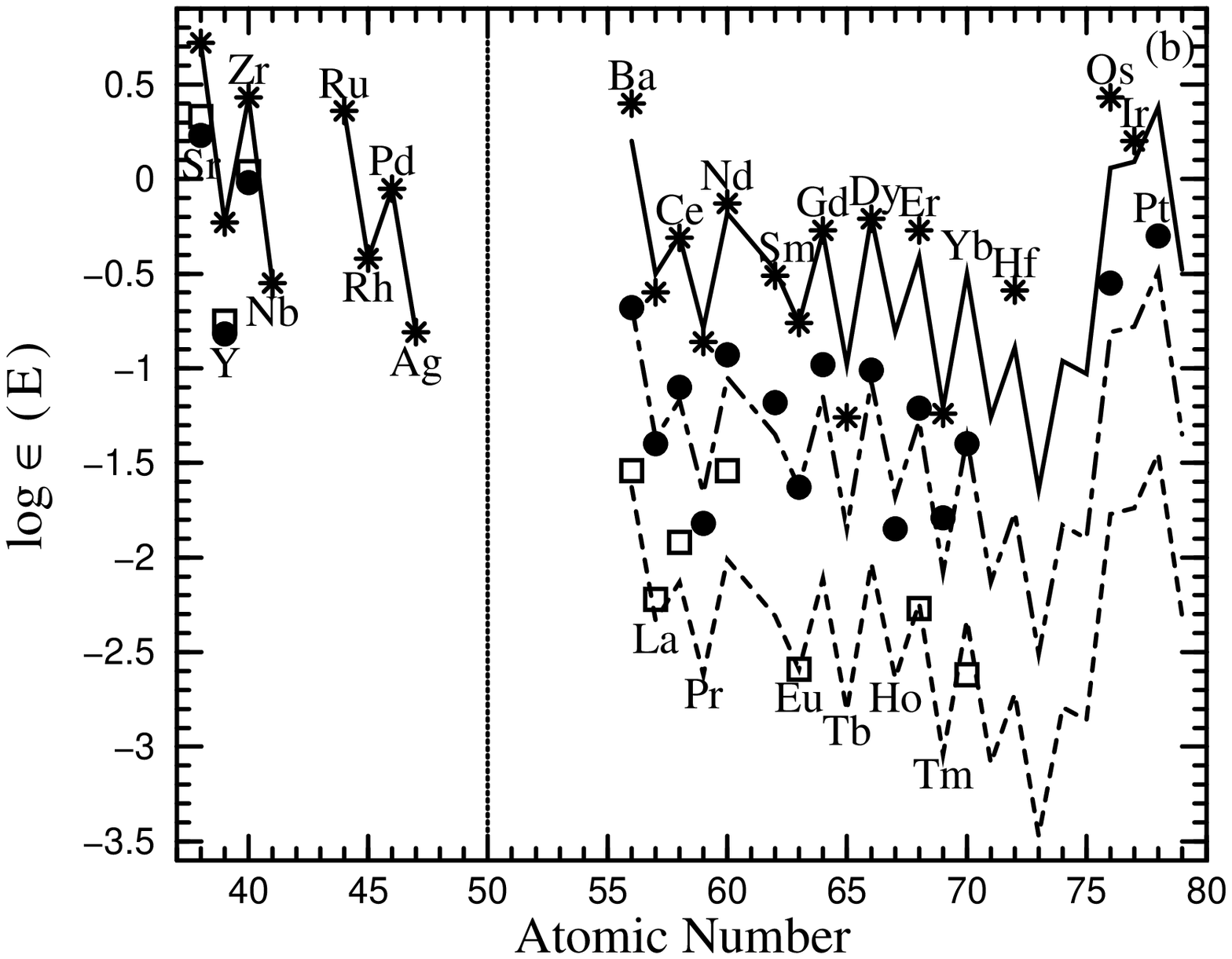}
\caption{Comparison of the observed abundances in CS 31082-001 (asterisks: 
[\protect\refcite{hi02}]), HD 115444 (filled circles), and HD 122563 
(squares: [\protect\refcite{we00}]). (a) The data on CS 31082-001 are 
connected with solid line segments as a guide. Missing segments mean 
incomplete data. The downward arrow at the asterisk for N indicates an 
upper limit. Note that the abundances of the elements from O to Ge are 
almost indistinguishable for the three stars. (b) The data on CS 31082-001 
to the left of the vertical line are again connected with solid line 
segments as a guide. In the region to the right of the vertical
line, the solid, dot-dashed, and dashed curves are the solar $r$-process
pattern translated to pass through the Eu data for CS 31082-001, HD 115444, 
and HD 122563, respectively. Note the close description of the data
by these curves. The shift between the solid and the dashed curves is 
$\sim 2$ dex.}
\end{center}
\end{figure}

Observations show that the late-time light curves of some core-collapse
supernovae were powered by decay of $^{56}$Ni to $^{56}$Fe 
(e.g. [\refcite{so02}]). These supernovae have progenitors of 
$>10\,M_\odot$, which develop Fe cores with extended shells before collapse. 
The supernova shock in this case produces $^{56}$Ni and lighter nuclei 
through explosive burning in the inner shells. This explosive burning and 
the hydrostatic burning in the outer shells during the presupernova 
evolution lead to production of the elements from O to Ni by core-collapse 
supernovae with progenitors of $>10\,M_\odot$. Such supernovae may be 
associated with the source responsible for mainly the light $r$-process 
nuclei (e.g. [\refcite{qw02}]).

\subsection{Tests for $r$-Process Sources}
The association of $r$-process sources with different progenitors of
core-collapse supernovae can be tested in a number of ways. First, it
needs to be shown that a substantial fraction of core-collapse supernovae
result from progenitors of $\sim 8$--$10\,M_\odot$ and AIC events. 
There are ongoing efforts to identify supernova progenitors by examining 
the archival images taken with the Hubble space telescope 
[\refcite{va03a,sm03}]. In one case, a progenitor of 
$\sim 8$--$10\,M_\odot$ has been demonstrated [\refcite{va03b,sm04}]. 
With sufficient statistics to be built up in the 
future, these efforts will be able to determine the fraction of 
core-collapse supernovae with low-mass progenitors. Identification of 
AIC events is more difficult as such events are not expected to have the
usual optical display associated with regular supernovae. A firm upper
limit on the rate of AIC events may be provided by super Kamiokande
and Sudbury Neutrino Observatory, which are capable of detecting the
neutrino signals from these events.

The strongest evidence so far for association of $r$-process 
nucleosynthesis with AIC events was provided by observations of the
metal-poor star HE 2148-1247 [\refcite{co03}]. The abundances of Fe and 
lower atomic numbers in this star are typically $\sim 10^{-2}$ times solar. 
However, the abundances of the neutron-capture elements Ba and above in the 
star are at the solar level (see Fig. 3). 
Such high neutron-capture abundances cannot 
represent the composition of the ISM from which HE 2148-1247 was formed.
Instead, they must have resulted from contamination of the surface of
this star. As abundances of Ba and above at the solar level must consist 
of contributions from both the $r$-process and the $s$-process (most of the 
solar Ba came from the $s$-process while essentially all the
solar Eu came from the $r$-process), the explanation for the observed 
abundances in HE 2148-1247 requires coordinated contamination by both an 
$r$-process source and an $s$-process source. This coordination may be
achieved as follows [\refcite{qw03}]. If HE 2148-1247 had a more massive 
binary companion of $\sim 3$--$8\,M_\odot$, this companion would have 
produced $s$-process elements during its evolution and contaminated the 
surface of HE 2148-1247 with these elements through mass transfer. The 
companion eventually became a white dwarf, which then collapsed into a 
neutron star [\refcite{no91}] after accreting some material back from 
HE 2148-1247. The ejecta from the AIC event would again contaminate the 
surface of this star, but this time with $r$-process elements. The 
extremely high contributions from both the $s$-process and the $r$-process 
to HE 2148-1247 can then be accounted for.

\begin{figure}[htb]
\begin{center}
\includegraphics[scale=0.4,angle=270]{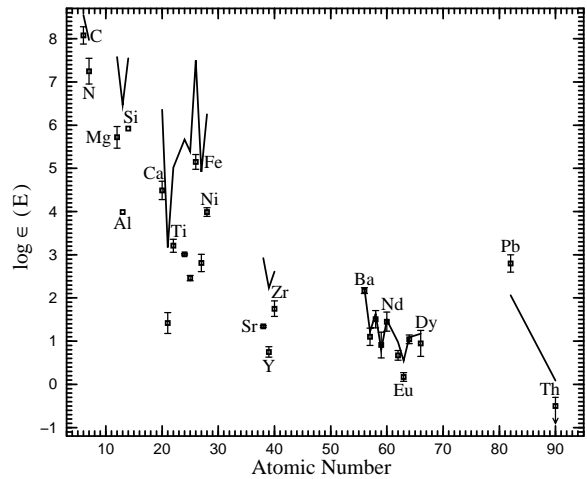}
\caption{The observed abundances in HE 2148-1247 (squares with 
error bars: [\protect\refcite{co03}]) compared with the solar abundances
(solid curve: [\protect\refcite{an89}]). The abundances of Fe and lower
atomic numbers in HE 2148-1247 are typically $\sim 10^{-2}$ times solar
but those of the elements Ba and above are at the solar level. Variations
in the radial velocity of HE 2148-1247 have been observed so this star is
in a binary [\protect\refcite{co03}].}
\end{center}
\end{figure}

The above AIC scenario has two possible outcomes. If the binary was 
disrupted by the AIC event, then a single metal-poor star with extremely 
high neutron-capture abundances would be produced. Indeed, a number of
such stars have been observed [\refcite{pr01}]. On the other hand, if the 
binary survives, then the metal-poor star with extremely high 
neutron-capture abundances has a neutron star companion today. The neutron 
star companion may be revealed through X-ray emission due to its accretion. 
A number of metal-poor stars with extremely high neutron-capture abundances, 
including HE 2148-1247, are known to be in binaries. Some loose upper limits 
on the X-ray emission due to possible neutron star companions of several 
stars were obtained from archival all-sky data [\refcite{sc03}]. It would be 
very interesting to see if dedicated X-ray searches can provide support for 
the AIC scenario.

The best tests for $r$-process production by core-collapse supernovae
would be direct observations of newly-synthesized $r$-process nuclei in
these events. One possibility is to examine the spectra of an event for
atomic lines of $r$-process elements. The lines of Ba were observed in
SN 1987A [\refcite{wi87}--\refcite{maz95}]. Combined with Fe production 
inferred from the light curve, this appears to indicate that SN 1987A 
produced both Fe and Ba. As mentioned in Sec. 2.1, Fe producing 
core-collapse supernovae may be responsible for mainly the light $r$-process
nuclei with $A\leq 130$. So the Ba observed in SN 1987A may
represent the heaviest $r$-process element produced
by such supernovae [\refcite{qw01}]. Clearly, optical 
observations of supernovae can provide very valuable information on the 
$r$-process. 

Perhaps the most direct test for $r$-process production is 
detection of $\gamma$-rays from the decay of the unstable progenitor nuclei 
that are initially produced by the $r$-process. A number of such nuclei have 
significant $\gamma$-ray fluxes for recent or future supernovae in the Galaxy
(e.g. [\refcite{qvw98,qvw99}]). The most interesting
nucleus is $^{126}$Sn with a lifetime of $\sim 10^5$ yr. If the Vela
supernova that occurred $\sim 10^4$ yr ago produced  $^{126}$Sn, the
$\gamma$-ray flux from the decay of this nucleus in the Vela supernova
remnant may be close to the detection limit of INTEGRAL [\refcite{qvw98}]. 
Hopefully, some information on $^{126}$Sn production by the Vela supernova 
will be provided by INTEGRAL in the near future.

\section{Yield Patterns and Role of neutrinos}
The extensive $r$-process patterns observed in the metal-poor 
stars CS 22892-052 [\refcite{sn00}] and CS 31082-001 [\refcite{hi02}]
are shown in Fig. 4. These two rather close patterns cover the elements 
from Sr to Cd with $A=88$--116 and the elements Ba and above with $A>130$.
Unfortunately, the elements Te and Xe with $A\sim 130$
cannot be observed in stellar spectra. On the other hand, Te and Xe
isotopes have been found in presolar diamonds. It is extremely important
to understand the Te and Xe patterns in these diamonds 
(e.g. [\refcite{ot96,ri98}]) as they may be the only data outside the solar 
system on $r$-process production in the region of $A\sim 130$.

\begin{figure}[htb]
\begin{center}
\includegraphics[scale=0.4,angle=270]{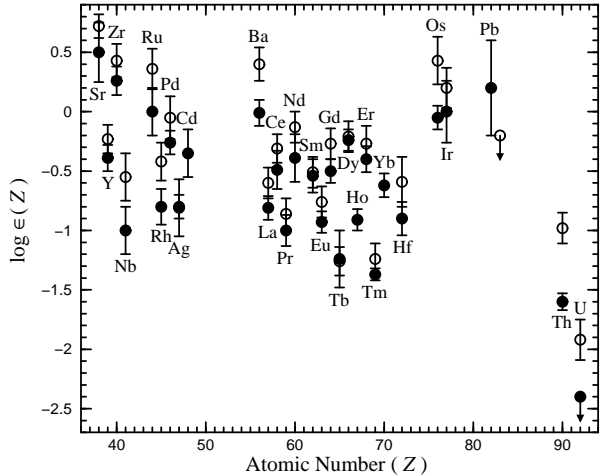}
\caption{The observed abundances in CS 22892-052 (filled circles: 
[\protect\refcite{sn00}]) compared with those in CS 31082-001
(open circles: [\protect\refcite{hi02}]). The open circle for Pb is 
shifted slightly for clarity. Downward arrows indicate upper limits.}
\end{center}
\end{figure}

According to the meteoritic data, at least some $r$-process events
produce $^{182}$Hf but not $^{129}$I. If these events were responsible
for the $r$-process patterns shown in Fig. 4, they must produce the
nuclei with $A=88$--116 and $A>130$ but skip those with $A\sim 130$.
A fission scenario was proposed in [\refcite{qian02}]
to achieve this. In this scenario, the $r$-process produces a freeze-out 
pattern covering $A>190$ with a peak at $A\sim 195$ and fission of 
progenitor nuclei during their decay towards stability produces the nuclei 
with $A<130$ and $A>130$ but very little of those with $A\sim 130$. 
In addition,
fission may be significantly enhanced through excitation of the progenitor 
nuclei by interaction with the intense neutrino flux in core-collapse 
supernovae [\refcite{qian02,ko04}]. The above fission scenario also
accounts for the difference in the Th/Eu ratio between CS 22892-052 and 
CS 31082-001 (see Fig. 4) that is difficult to explain by the possible
difference in the stellar age [\refcite{qian02}]. In this scenario,
Th represents the progenitor nuclei surviving fission whereas Eu represents
the nuclei produced by fission. The Th/Eu ratio would then depend on the 
fraction of the progenitor nuclei undergoing fission, which would in turn
depend on, for example, the extent of neutrino interaction in a specific 
$r$-process event.

Neutrino interaction can also result in emission of neutrons 
[\refcite{qi97,la01}]. In fact, it was shown that the solar $r$-process 
abundances of the nuclei with $A=183$--187 can be completely
accounted for by neutrino-induced neutron emission from the progenitor
nuclei in the peak at $A\sim 195$ [\refcite{qi97,ha97}]. Clearly, to test 
neutrino-induced nucleosynthesis associated with the $r$-process requires 
a lot of nuclear physics input, such as branching ratios of fission and 
neutron emission for excited neutron-rich nuclei and the associated fission
yields. Perhaps some of this input could be studied at future rare
isotope accelerator facilities.

\begin{figure}[htb]
\begin{center}
\includegraphics[scale=0.4]{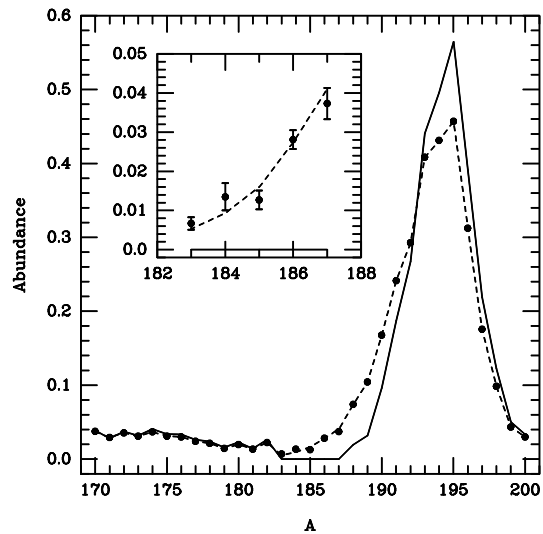}
\caption{Effects of neutrino-induced neutron emission for the region
near the abundance peak at $A\sim 195$ [\protect\refcite{qi97,ha97}]. The 
abundances before and after neutrino-induced neutron emission (following 
freeze-out of the $r$-process) are given by the solid and dashed curves, 
respectively. The filled circles (some with error bars) give the solar 
$r$-process abundances. The region with $A=183$--187 is highlighted in the 
inset.}
\end{center}
\end{figure}

\end{document}